%% file: TOP-13-015_temp.tex
\pdfoutput=1

\documentclass[11pt,twoside,a4paper,cmspaper,final,collab]{cms-tdr}
\def\svnVersion{343483}\def\cmsCernNoTag{CERN-PH-EP/2015-289}\def\cmsCernDate{\today}\def\cmsMessage{Published in Physics Letters B as \href{http://dx.doi.org/10.1106/j.physletb.2016.05.005}{\doi{10.1106/j.physletb.2016.05.005.}}}

\begin{document}\cmsNoteHeader{TOP-13-015}

\hyphenation{had-ron-i-za-tion}
\hyphenation{cal-or-i-me-ter}
\hyphenation{de-vices}
\RCS$HeadURL: svn+ssh://svn.cern.ch/reps/tdr2/papers/TOP-13-015/trunk/TOP-13-015.tex $
\RCS$Id: TOP-13-015.tex 343422 2016-05-19 07:36:41Z ksbeerna $
\newlength\cmsFigWidth
\ifthenelse{\boolean{cms@external}}{\setlength\cmsFigWidth{0.85\columnwidth}}{\setlength\cmsFigWidth{0.4\textwidth}}
\ifthenelse{\boolean{cms@external}}{\providecommand{\cmsLeft}{top\xspace}}{\providecommand{\cmsLeft}{left\xspace}}
\ifthenelse{\boolean{cms@external}}{\providecommand{\cmsRight}{bottom\xspace}}{\providecommand{\cmsRight}{right\xspace}}
\providecommand{\NA}{---\xspace}

\newcolumntype{x}{D{,}{\, \pm \,}{5,2}}
\newcolumntype{X}{D{,}{\, \phantom{\pm} \,}{5,2}}
\newcolumntype{.}{D{.}{.}{2.3}}

\cmsNoteHeader{TOP-13-015}
\title{ Measurement of spin correlations in \texorpdfstring{$\ttbar$}{t-tbar} production using the matrix element method in the muon+jets final state in pp collisions at \texorpdfstring{$\sqrt{s} = 8\TeV$}{sqrt(s) = 8 TeV}}

\date{\today}

\abstract{
The consistency of the spin correlation strength in top quark pair production with the standard model  (SM) prediction is tested in the muon+jets final state.  The events are selected from pp collisions, collected by the CMS detector, at a centre-of-mass energy of 8\TeV,  corresponding to an integrated luminosity of 19.7\fbinv. The data are compared with the expectation for the spin correlation predicted by the SM and with the expectation of no correlation. Using a template fit method, the fraction of events that show SM spin correlations is  measured to be $0.72\pm 0.08 \stat {}^{+0.15}_{-0.13}\syst$, representing the most precise measurement of this quantity in the muon+jets final state to date.
}

\hypersetup{%
pdfauthor={CMS Collaboration},%
pdftitle={Measurement of spin correlations in ttbar production using the matrix element method in the muon+jets final state in pp collisions at sqrt(s) = 8 TeV},%
pdfsubject={CMS},%
pdfkeywords={CMS, LHC, top quark, matrix element method, spin}}

\maketitle
\section{Introduction}
\label{sec:Introduction}
At the CERN LHC top quarks are predominantly produced in pairs ($\ttbar$), mainly via gluon fusion, with each top quark decaying almost 100\% of the time
into a W boson and a b quark. The final states can be categorised as dilepton,
where both W's decay into a lepton and a neutrino, hadronic, where both
W's decay into quarks, and lepton+jets otherwise. The W decay into a tau lepton and neutrino is only considered leptonic if the $\tau$ decays include a muon or electron.

In quantum chromodynamics (QCD), the quark spins in heavy quark
production are correlated.
Since the lifetime of top quarks is smaller than
the hadronisation timescale ($1/\Lambda_{\mathrm{QCD}}$), which in turn is smaller than the spin decorrelation timescale $m_{\PQt}/\Lambda_{\mathrm{QCD}}^2\sim3\times10^{-21}\unit{s}$, the top quarks decay before their spins decorrelate.
This spin correlation is therefore propagated to the top quark decay products and one can infer the $\ttbar$ spin correlation strength $A$ by studying the angular correlations between the decay products, where
\begin{equation}
A=\frac{(N_{\uparrow\uparrow}+N_{\downarrow\downarrow}) - (N_{\uparrow\downarrow} + N_{\downarrow\uparrow})}{(N_{\uparrow\uparrow}+N_{\downarrow\downarrow})+ (N_{\uparrow\downarrow} + N_{\downarrow\uparrow})}
\end{equation}
is the asymmetry between the number of $\ttbar$ pairs with aligned and antialigned spins. The value of $A$ depends on the spin
quantization axis chosen and on the production modes.

Given the high centre-of-mass energy at the LHC, the helicity basis is used where the spin quantization axis is defined as the top quark or antiquark direction in the $\ttbar$ rest frame. The corresponding value of the spin correlation strength in the helicity basis is referred to as $A_\text{hel}$. Since the spin correlation strength is precisely, but non-trivially, predicted by the standard model (SM) an accurate measurement of this variable tests various aspects of the SM, including the strength
of the QCD coupling and the relative contribution of $\ttbar$ production modes, although new
physics can influence the spin correlation strength~\cite{krohn2011,brenreuther2013}.

Tevatron experiments made measurements of the $\ttbar$ spin correlation strength using template fits to the angular
distributions of the top quark decay products and extracting the fraction of $\ttbar$ events with the SM prediction of spin correlation $f$
defined as
\begin{equation}
f = \frac{N^{\ttbar}_\mathrm{SM}}{N^{\ttbar}_\mathrm{SM} + N^{\ttbar}_\text{uncor}},
\label{eq.Deff}
\end{equation}
where $N^{\ttbar}_\mathrm{SM}$ is the number of SM $\ttbar$ events, whereas $N^{\ttbar}_\text{uncor}$
represents the number of events with uncorrelated $\ttbar$. The top quark and antiquark in the uncorrelated $\ttbar$ events decay spherically.
The assumption is that there are only SM and uncorrelated $\ttbar$ events, with a fraction of $(1-f)$ of uncorrelated $\ttbar$ events. The physical range of this parameter $f$ is restricted to [0, 1], with $f = 1$ for a sample of $\ttbar$ events produced by the SM. However, quite often an unconstrained template fit is performed, allowing for non-physical values of this parameter. The CDF Collaboration extracted the fraction $f$ of events with the SM prediction of spin correlation
using the lepton+jets final state~\cite{CDF2011}
and the \DZERO Collaboration extracted this fraction using the dilepton final states~\cite{D002000,D02011}.
The \DZERO Collaboration also made a spin correlation measurement using the matrix
element method (MEM)~\cite{D0MEM12011} in the dilepton final state and found direct evidence of $\ttbar$ spin correlation
by combining the measurements using MEM in the dilepton and lepton+jets final states~\cite{D0MEM2012}.
The combined measurement yielded $f = 0.85 \pm 0.29\,\text{(stat + syst)}$ using a data sample of $\Pp\PAp$ collisions at $\sqrt{s} = 1.96\TeV$, corresponding to an integrated luminosity of 5.3\fbinv.

At the LHC, the ATLAS Collaboration has reported observation of spin correlations in top quark pair production~\cite{ATLASResultOld}. In the most recent measurement by the ATLAS Collaboration, the spin correlation measurement was performed using template fits to the distribution of the difference in azimuthal angle between the two oppositely charged leptons in the dilepton final state. This measurement at $\sqrt{s} = 8\TeV$, using 20.3\fbinv of integrated luminosity, resulted in $f = 1.20 \pm 0.05\stat \pm 0.13\syst$~\cite{AtlasResult2012}. Another result by ATLAS in the dilepton channel has been reported in~\cite{1510.07478}.
The only measurement in the lepton+jets final state at the LHC so far was made by the ATLAS Collaboration using the opening angle distributions between the decay products of the top quark and antiquark~\cite{ATLASnewSC}, giving $f  =  1.12 \pm 0.11\stat \pm 0.22\syst$ at $\sqrt{s} = 7\TeV$, using 4.6\fbinv of integrated luminosity.

Here, a measurement of the top quark spin correlations in events characterised by the presence of a muon and jets ($\mu$+jets) is described using a MEM at $\sqrt{s} = 8\TeV$ with 19.7\fbinv of integrated luminosity. Events with a muon coming from a $\tau$ decay are not considered as part of the signal. In this analysis, the traditional
discrete hypotheses are investigated: SM and uncorrelated $\ttbar$ production and decay. In the MEM, the likelihood of an observed event to be produced by a
given theoretical model is calculated. The likelihood ratio of the sample allows to distinguish between the two hypotheses. In addition,
the distribution of event likelihood ratios is used in a template fit to extract the fraction $f$ of events with the SM prediction of spin correlation.

The rest of this Letter is organised as follows. In Section~\ref{sec:Detector Description}, a description of the apparatus used in this measurement, the CMS detector, is given. Following, in Section~\ref{sec:Signal and Background Modeling}, a description of the simulation samples used in this analysis is given. The event selection and reconstruction procedure of the physics objects in an event are given
in Section~\ref{sec:Object Reconstruction and Event Selection}. In Section~\ref{sec:MEM}, the MEM is briefly explained. Section~\ref{sec:Hypothesis Testing} describes the first part of this analysis, the hypothesis-testing procedure, followed by the extraction of the variable $f$ with a template fit in Section~\ref{sec:Extractionf}. The sources of systematic uncertainties are discussed in Section~\ref{sec:Systematic Uncertainties}. A description on the treatment of these uncertainties in both parts of the analysis and the results are given in Section~\ref{sec:Results}. Finally, a summary of the analysis is presented in Section~\ref{sec:Summary}.

\section{The CMS detector}
\label{sec:Detector Description}
The central feature of the CMS apparatus~\cite{Chatrchyan:2008zzk} is a 3.8\unit{T} superconducting solenoid of 6\unit{m} internal diameter.
The silicon pixel and strip tracker used for
measuring charged-particle trajectories, a lead tungstate
crystal electromagnetic calorimeter (ECAL), and the brass and scintillator hadron calorimeter (HCAL) are located within
the superconducting solenoid volume. The calorimeters, ECAL and HCAL, both of which consist of a barrel and two endcap
sections, surround the silicon tracking volume. Forward calorimetry extends the coverage provided by the barrel and endcap detectors to a pseudorapidity of $\abs{\eta} = 5$.

Muons are measured using the tracker and the muon system that consists of gas-ionization detectors embedded in the steel flux-return yoke
outside the solenoid. Muons are measured in the range $\abs{\eta}< 2.4$, using three detector technologies: drift tubes,
cathode strip chambers, and resistive-plate chambers. Matching muons to tracks measured in the silicon tracker results in
a relative transverse momentum (\pt) resolution of 1.3--2.0\% in the barrel and better than
6\% in the endcaps for muons with $20 <\pt < 100\GeV$. The \pt resolution in the barrel is better than 10\% for muons with \pt up to 1\TeV~\cite{Chatrchyan:2012xi}.

The first level of the CMS trigger system, composed of custom hardware processors, uses information from the calorimeters
and muon detectors to select the most interesting events. The high-level trigger
processor farm further decreases the event rate from around 100\unit{kHz} to around 400\unit{Hz}, before data storage.

A more detailed
description of the CMS detector, together with a definition of the coordinate system used, can be found in Ref.~\cite{Chatrchyan:2008zzk}.

\section{Signal and background modeling}
\label{sec:Signal and Background Modeling}
The signal processes ($\ttbar$ events in the $\mu$+jets final state, SM and uncorrelated) as well as other $\ttbar$ decay channels (SM and uncorrelated)
are simulated on the basis of a next-to-leading-order (NLO)
calculation using the generator \MCATNLO v3.41~\cite{Frixione} with a top quark mass of 172.5\GeV.
Parton showering is simulated using \HERWIG 6.520~\cite{Herwig} and the default \HERWIG 6 underlying event tune was used.
The NLO parton distribution function (PDF) set used is \textsc{cteq6m}~\cite{Lai2010}.
The background samples of W+jets and Z/$\gamma$*+jets processes are generated using \MADGRAPH 5.1.3.30~\cite{MadGraph5_2011},
\PYTHIA~6.426, and \TAUOLA~v27.121.5~\cite{Was2000}.
The backgrounds from single top quark processes are generated using \POWHEG~v1~\cite{Nason:2004rx,Frixione:2007vw,Alioli:2010xd} and \TAUOLA~\cite{Davidson:2010rw}.
The Z2* underlying event tune is used. The most recent \PYTHIA Z2* tune is derived from the Z1 tune~\cite{Field:2010bc}, which uses the CTEQ5L parton distributions set, whereas Z2* adopts CTEQ6L~\cite{1126-6708-2002-07-012}.
The generated events are processed through the CMS detector simulation based on \GEANTfour~\cite{GEANT4} and event reconstruction.
To estimate the size of the effect of the top quark mass and factorisation and renormalisation scale uncertainties, \MCATNLO samples with varied top quark mass and scales are used.
The signal event yields are scaled to match the predicted top quark pair production cross section in proton-proton
collisions at $\sqrt{s} = 8\TeV$, which is $\sigma^\mathrm{NNLO+NNLL}_{\ttbar} =  245.8^{+6.2}_{-8.4}\,\text{(scales)} {}^{+6.2}_{-6.4}\,\text{(PDF)}\unit{pb}$ for a top quark mass equal to the world average of 173.3\GeV~\cite{ATLAS:2014wva}, computed with next-to-next-to-leading-order (NNLO) QCD corrections and next-to-next-to-leading-logarithmic (NNLL) resummation accuracy~\cite{Czakon:2013}. The simulated samples for the background processes are normalized using cross section calculations, generally at NLO accuracy~\cite{Czakon:2013}. Where necessary, systematically varied cross sections have been used for the normalization. The simulation is corrected to the pileup conditions seen in the data. Pileup refers to
the additional proton-proton interactions recorded simultaneously from the same bunch crossing. During 2012 data taking, there were on average 20 interactions per bunch crossing.

\section{Event reconstruction and selection}
\label{sec:Object Reconstruction and Event Selection}
The event selection has been optimized to identify $\ttbar$ events in the $\mu$+jets final state.
A single-muon trigger with a muon \pt threshold of 24\GeV and a restriction on the pseudorapidity $\abs{\eta} < 2.1$ is used to collect the data samples.
Isolation and identification criteria are applied at the trigger level to achieve manageable rates with minimal loss of efficiency.

The physics objects used in this analysis are reconstructed with the CMS particle-flow (PF) algorithm~\cite{CMS-PAS-PFT-09-001, CMS-PAS-PFT-10-001}. The PF algorithm reconstructs and identifies each individual particle in an event using combined
information from all CMS subdetectors. The energy of photons is directly obtained from the ECAL measurement. The energy of electrons is determined from a combination of
the electron momentum measured at
the primary interaction vertex by the tracker, the energy of the matched ECAL cluster, and the total energy of
the associated bremsstrahlung photons. The momentum of muons is obtained from
the curvature of the track associated to the muon. The energy of charged hadrons is determined from a combination of their momenta
measured in the tracker and the matching energy deposits in the ECAL and HCAL, corrected for the
calorimeter response to hadronic showers. Finally, the energy of neutral hadrons is obtained from the
corresponding corrected ECAL and HCAL energy.

The reconstructed
muon candidates are required to have $\pt > 26\GeV$ and $\abs{\eta} < 2.1$, as to be in a region where the trigger is fully efficient. The track associated to the muon candidate is
required to have a minimum number of hits in the silicon tracker, to be consistent with the primary vertex, and to have
a high-quality fit which combines a track in the tracker and a minimum number of hits in the muon detectors into one track. For each muon candidate, a PF-based
relative isolation is calculated, corrected for pileup effects on an event-by-event basis. The transverse momenta of all reconstructed particle candidates (excluding the muon itself) are summed in a cone of size $\Delta R < 0.4$ around the muon direction, with $\Delta R = \sqrt{\smash[b]{(\Delta\eta)^{2} + (\Delta\phi)^{2}}}$ where $\phi$ is the azimuthal angle expressed in radians. The pileup contribution in this scalar sum is corrected for by summing only over the charged particles associated to the event vertex in the charged particle contribution, and subtracting the average energy due to pileup in the neutral particle contribution.
After subtraction of the pileup contribution, the scalar sum is required to be smaller than 12\% of the muon \pt. It is required that exactly one of these well-identified muon candidates is present in the event. In addition, a looser selection on muons is applied which requires a relative isolation of less than 20\% of the muon \pt, a selection of $\pt > 10\GeV$ and $\abs{\eta} < 2.5$. Events with additional muons passing
looser identification criteria, as well as events with an electron are discarded. Events selected from other $\ttbar$ final states are denoted as ``$\ttbar$ other'' and consist of roughly 70\% $\ttbar$ events in the dilepton final state and 30\% events in the $\tau$+jets final state.

For each event, hadronic jets are clustered from the reconstructed particle-flow particles with the anti-\kt algorithm~\cite{Cacciari:2008gp, Cacciari:2011ma}, with a distance parameter of 0.5.
The jet momentum is determined as the vector sum of all particle momenta in the jet, which has been determined from simulation to be within
5\% to 10\% of the true momentum over the whole \pt spectrum and detector acceptance. Contributions from pileup are taken into account by
an offset correction to the jet energies. Jet energy scale corrections (JES) up to particle-level
are derived from simulation, and are confirmed with in-situ measurements of the energy balance in dijet and photon+jet events.
The jet energy resolution (JER) in simulation is corrected to match the resolution observed in data. Additional selection
criteria are applied to each event to remove spurious jet-like features originating from isolated uncharacteristic
noise patterns in certain HCAL regions~\cite{Chatrchyan:2010ef} and in the silicon avalanche photodiodes used in the ECAL barrel detector.
The first three jets leading in \pt are required to have a \pt of at
least 30\GeV, the fourth leading jet of at least 25\GeV and the remaining jets at least
20\GeV. At least two selected jets should be identified as coming from the decay of B-hadrons, based on the combined secondary vertex (CSV)
algorithm with medium working point (CSVM)~\cite{CSVM2012}. The CSV
algorithm makes use of secondary vertices, when available, combined with track-based b-lifetime information. As the tracker coverage is limited to $\abs{\eta} < 2.4$, all selected jets (both tagged and untagged) are restricted to this pseudorapidity range. The missing transverse momentum vector \ptvecmiss is defined as the projection on the plane perpendicular to the beams of the negative vector sum of the momenta of all reconstructed particles in an event. Its magnitude is referred to as \ETmiss.
To reduce the effect of Final State Radiation (FSR), while not statistically limiting the analysis, we restrict the data set to events with four or five selected jets.
To ensure that the selected jets in the event describe the $\ttbar$ kinematic quantities, we reject events if they have additional forward jets in the region of $2.4 < \abs{\eta} < 4.7$ and these have $\pt > 50\GeV$.

\begin{figure}[ht!]
\centering
\includegraphics[width=0.45\textwidth]{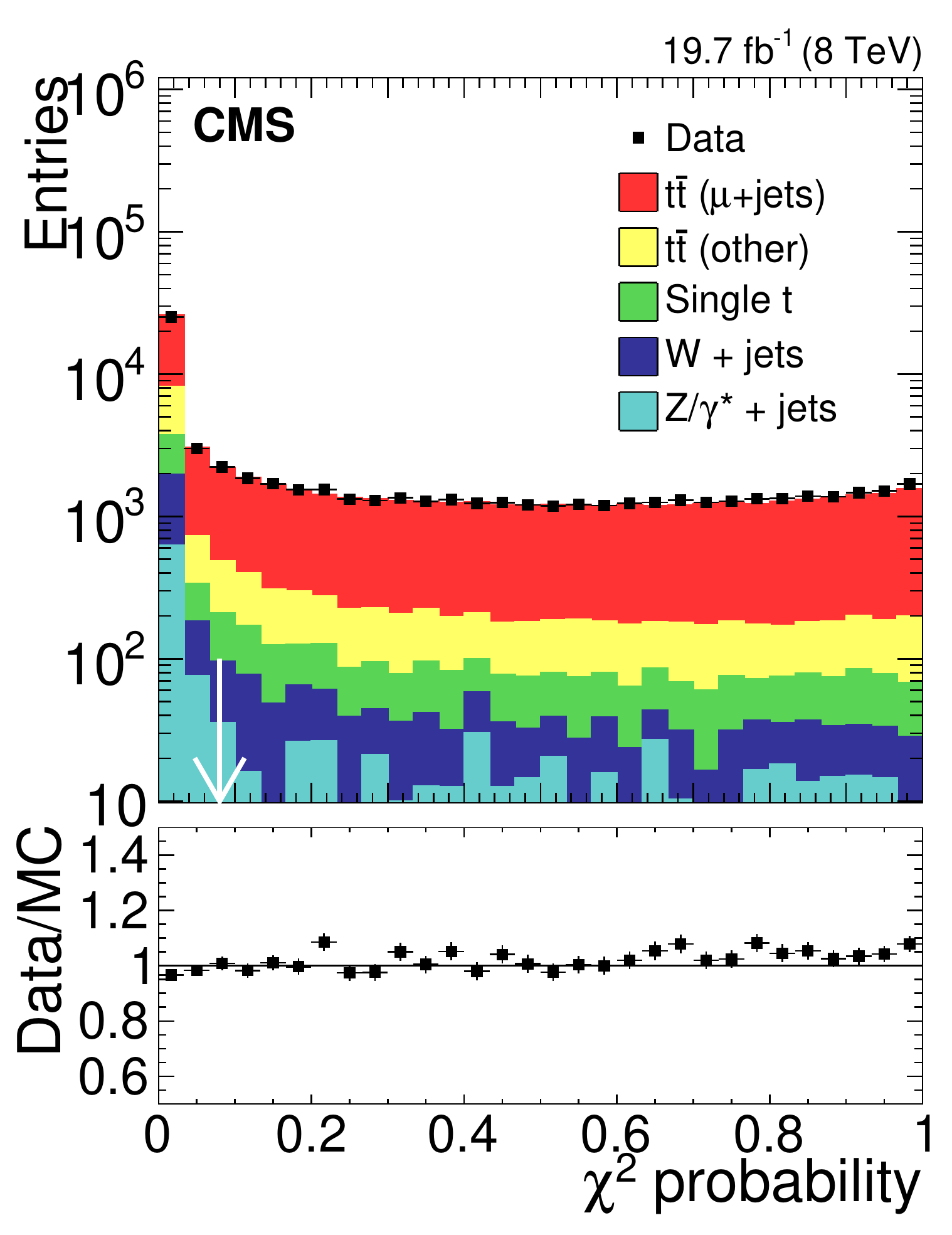}
\caption{The $\chi^{2}$ probability distribution of the selected solutions of the kinematic fit in the $\mu$+jets channel, showing a shape comparison between data and simulation including the statistical uncertainties. The relative contributions in simulation are calculated using the theoretical cross sections with the total yield normalised to data. For the analysis, we only consider events with a probability larger than 0.08, as indicated by the arrow.}
\label{fig:Chi2}
\end{figure}

To further increase the quality of the event selection and reduce the background contribution, we use a kinematic fitter, {\sc HitFit}~\cite{PhysRevD.58.052001},
designed to reconstruct the kinematic quantities of the $\ttbar$ system in the lepton+jets final state. The kinematic quantities observed in the event are varied within the detector
resolution to satisfy some predefined constraints, i.e.\ the reconstructed hadronically decaying W boson mass is required to be consistent with 80.4\GeV and
the reconstructed top quark and antiquark masses are required to be equal. The {\sc HitFit} algorithm tries every jet-quark permutation and the solution with the highest goodness-of-fit (or equivalently, lowest $\chi^{2}/\mathrm{ndof}$ with ndof being the number of degrees of freedom)
is chosen as the best estimate of the correct jet-quark permutation.
We do not rely on \textsc{HitFit} to estimate the jet-quark permutation correctly, however, \textsc{HitFit} is used to decide which four jets in the event to use in the reconstruction of the $\ttbar$ final state in five-jet events. It is required that two of the jets selected by \textsc{HitFit} are identified as originating from B-hadrons. The selection of the jets in the event could be done with simpler methods, e.g. selecting the highest-\pt jets, but \textsc{HitFit} offers the possibility to apply additional quality criteria. In order to reduce the background fraction and the fraction of mismodeled events, we only select events with a \textsc{HitFit} $\chi^{2}/\mathrm{ndof} < 5 $ or, equivalently, with the fit probability larger than 0.08.
The value of the $\chi^{2}/\mathrm{ndof}$-selection is chosen to maximize the separation
power defined by Eq.~(\ref{eq.SepPower}) in Section~\ref{sec:Hypothesis Testing}. Mismodeled events can be due to the inclusion of radiated jets in the $\ttbar$ reconstruction or events with poorly reconstructed jet quantities. The $\chi^{2}$ probability distribution is shown for data and simulation in Fig.~\ref{fig:Chi2}, where the relative contributions of the simulation are determined from the theoretical cross sections.

The event yield after the full event selection is displayed in Table~\ref{table:EvYield}. The contributions are estimated from simulation and normalised to the observed luminosity using theoretical cross sections. The selection efficiency for the SM and uncorrelated signal samples are very similar so that the event selection does not bias the data towards one hypothesis. The background contribution due to multijet processes has been estimated from simulation and is found to be negligible.

\begin{table}[ht!]
\topcaption{Event yield after event selection, with the statistical uncertainties. The contributions from various physics processes are given, with a comparison between the data and the total simulation at the bottom.}
\centering
\begin{tabular}{l x}
\hline
Process & \multicolumn{1}{c}{Yield} \\
\hline
W+jets & 722,20\\
Z/$\gamma$*+jets & 139,18\\
$\PQt, \PAQt$ ($s$ channel) & 41,3\\
$\PQt, \PAQt$ ($t$ channel) & 314,10\\
$\PQt, \PAQt$ (tW) & 935,20\\
$\ttbar$ other & 3\, 896,24\\
$\ttbar$ $\mu$+jets & 31\, 992,69\\
\hline
Total simulation & 38\, 039,81\\
Data & \multicolumn{1}{X}{37\, 775}\\
\hline
\end{tabular}
\label{table:EvYield}
\end{table}

\section{Matrix element method}
\label{sec:MEM}
The matrix element method ~\cite{kondo1988,kondo1991,dalitz1992,d0mem2004} is a technique that directly relates theory with experimental events.
The compatibility of the data recorded with the leading-order (LO) matrix element (ME) of a certain process is evaluated. The probability that an event is produced by this process is calculated using the full kinematic information in the event.

The probability $P(x_i|H)$ to observe an event $i$ with kinematic properties $x$ for a certain hypothesis $H$ is given by:
\begin{equation}
\ifthenelse{\boolean{cms@external}}
{
\begin{split}
P(x_i|H) = \frac{1}{\sigma_\mathrm{obs}(H)} & \int   f_\mathrm{PDF}(q_1)f_\mathrm{PDF}(q_2) \rd q_1 \rd q_2 \\ & \frac{(2\pi)^4\left|M(y,H)\right|^2}{q_1q_2s}W(x_i,y) \rd \Phi_6.
\end{split}
}
{
P(x_i|H) = \frac{1}{\sigma_\mathrm{obs}(H)}\int f_\mathrm{PDF}(q_1)f_\mathrm{PDF}(q_2) \rd q_1 \rd q_2\frac{(2\pi)^4\left|M(y,H)\right|^2}{q_1q_2s}W(x_i,y) \rd \Phi_6.
}
\label{eq.EvProb}
\end{equation}

The given probability is equivalent to an event likelihood. In this equation, $q_1$ and $q_2$ represent the parton energy fractions in the collision, $f_\mathrm{PDF}(q_{1})$ and $f_\mathrm{PDF}(q_{2})$ are the PDFs, $s$ is the centre-of-mass energy squared of the colliding protons,
and $\rd \Phi_6$ represents the phase space volume element.
The transfer function, $W(x,y)$,  relates observed kinematic quantities $x$ with parton-level quantities $y$.
For every $y$, the transfer function is normalised to unity by integrating over all possible values
of $x$.
The LO ME is represented by $M(y,H)$, where $H$ denotes the hypothesis used.
The $\ttbar$ spin correlation strength is not a parameter of the SM Lagrangian, therefore $H$ is not a continuous parameter. The MEs $M(y,H_\mathrm{cor})$ and $M(y,H_\text{uncor})$ both describe $\ttbar$ production and subsequent decay in the $\mu$+jets channel valid for both on- and off-shell top quarks. In this analysis, the hypotheses are either the SM ($H_\mathrm{cor}$), giving rise to a finite value of the spin correlation strength $A$ (as discussed in Section~\ref{sec:Introduction}) or the spin-uncorrelated hypothesis with $A = 0$ ($H_\text{uncor}$). Finally, $\sigma_\mathrm{obs}(H)$ represents the observed $\ttbar$ cross section of the hypothesis, which ensures that the probability is normalised.
The quantity $\sigma_\mathrm{obs}(H)$ consists of the product of the production cross section $\sigma$, which is identical for our considered hypotheses, and the overall selection efficiency $\epsilon(H)$. The selection efficiency for events from both hypotheses are very similar, with an efficiency of $\epsilon(SM) = 0.0448 \pm 0.0001\stat $ for the SM $t\overline{t}$ signal hypothesis, and $\epsilon(uncor) = 0.0458 \pm 0.0001\stat $ for the uncorrelated signal hypothesis, which causes acceptance effects to nearly cancel in the likelihood ratio.
The likelihood calculation is performed
using \textsc{MadWeight}~\cite{Artoisenet:2010cn}, in the \MADGRAPH5 framework~\cite{MadGraph5_2011}. Since, in our convention, the likelihood
for a single event is represented by $P(x_i | H)$, the likelihood of a sample with $n$ events is then
\begin{equation}
L(x_1,\ldots,x_n|H)=\Pi_{i=1}^nP(x_{i}|H).
\label{eq.SampleLik}
\end{equation}

\begin{figure}[ht!]
\centering
\includegraphics[width=0.45\textwidth]{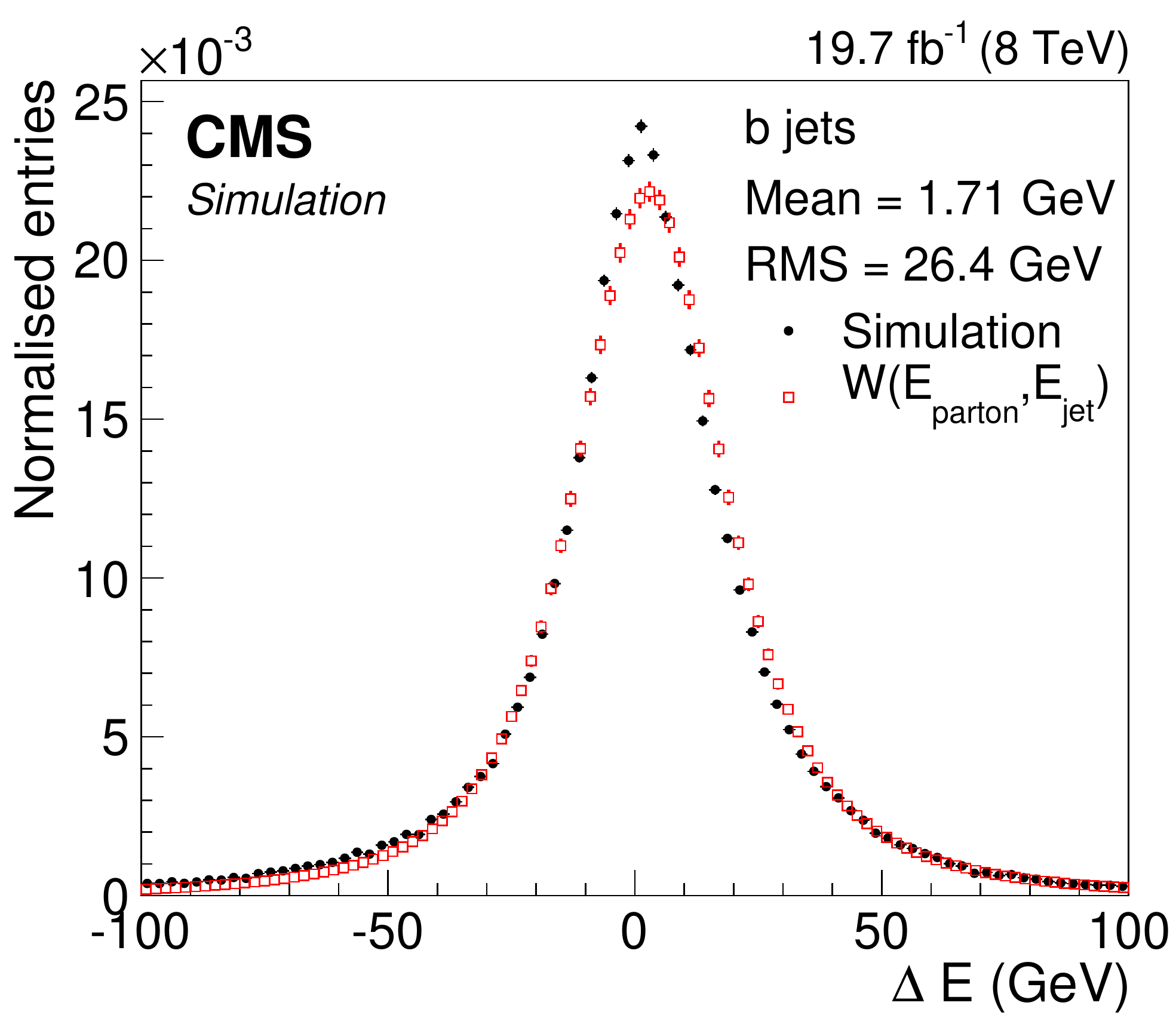}
\includegraphics[width=0.45\textwidth]{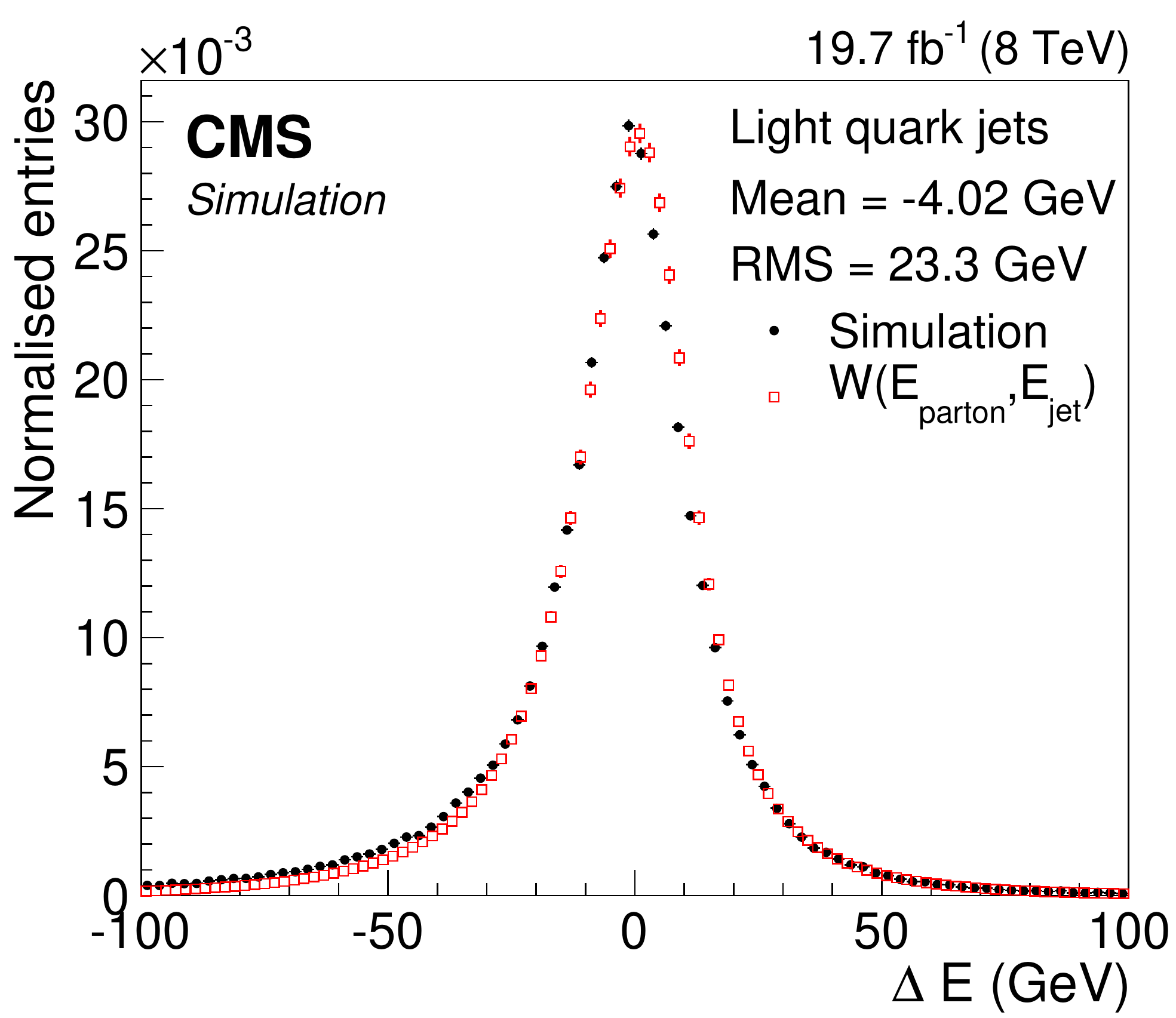}
\caption{$\Delta E$ distributions based on the values obtained from simulation (circles) compared to the $\Delta E$ distribution obtained by folding the $E_\text{parton}$ spectrum of matched partons with the transfer function (squares) summed over all values of $E_\text{parton}$ and $\abs{\eta_\text{parton}}$. The mean and RMS shown on the plots are obtained from simulation. The figure is shown for b quark jets (\cmsLeft) and for light quark jets (\cmsRight).}
\label{fig:TF}
\end{figure}

The transfer function of a given interacting particle depends on the
specifics of the detector. In this analysis, the transfer function is used to correct the jet kinematic quantities. The reconstructed jet energy information, corrected for JES and JER, is mapped onto
parton-level quantities by integrating over the parton energy within the transfer function resolution during the likelihood calculation. All other kinematic quantities (such as angular information or lepton quantities) are unmodified by the transfer function as these are measured with sufficient accuracy with the CMS detector to describe a final state that does not include a dilepton resonance. The description of these variables with a Dirac delta function speeds up the integration. The \ETmiss is also described with a Dirac delta function and is only used to correct the kinematic quantities of the event for the transverse Lorentz boost. The event
transfer function is the product of the object transfer functions, assuming no correlation between the reconstructed
objects. The jet energy
transfer function is determined from $\ttbar$ simulation to which the JES and JER corrections have been applied. For each jet in the simulation, unambiguously matched to a parton with
$\Delta R\mathrm{(jet,parton)} < 0.3$, the $E_\mathrm{jet}$ and $E_\text{parton}$ are compared (separately for jets matched to b and light-flavour partons).
The $E_\mathrm{jet}$ distribution is fitted with a Gaussian function, where the Gaussian mean and width depend on $E_\text{parton}$ and are given by $\mu(E_\mathrm{jet}) = m_{0}(\eta_\text{parton}) + m_{1}(\eta_\text{parton}) E_\text{parton}$ and $\sigma (E_\mathrm{jet}) =\sigma_{0}(\eta_\text{parton}) + \sigma_{1}(\eta_\text{parton})E_\text{parton} + \sigma_{2}(\eta_\text{parton})\sqrt{ \smash[b]{ E_\text{parton}}}$ respectively. The fit of the
$E_\mathrm{jet}$ distribution is converted to a single Gaussian transfer function, which is a function of the variable $\Delta E = E_\text{parton} - E_\mathrm{jet}$ and the parameters are a function of $E_\text{parton}$. The transfer function, which is determined in the full kinematic phase space, is given by
\begin{equation}
\ifthenelse{\boolean{cms@external}}
{
\begin{split}
W( & E_\text{parton}, \, E_\mathrm{jet}) = \\
& \ \ \  \ \frac{1}{\sqrt{2\pi} \left(\, \sigma_{0} + \sigma_{1}E_\text{parton} + \sigma_{2}\sqrt{\smash[b]{E_\text{parton}}} \, \right)} \\
&  \times\  \exp{\left[ -\frac{1}{2}\left(\frac{\Delta E + m_{0} + m_{1}E_\text{parton}}{\sigma_{0} + \sigma_{1}E_\text{parton} + \sigma_{2}\sqrt{\smash[b]{E_\text{parton}}}} \right)^{2} \right]},
\end{split}
}
{
\begin{split}
W(  E_\text{parton}, \, E_\mathrm{jet}) & =
 \ \frac{1}{\sqrt{2\pi} \left(\, \sigma_{0} + \sigma_{1}E_\text{parton} + \sigma_{2}\sqrt{\smash[b]{E_\text{parton}}} \, \right)} \\
&  \times\  \exp{\left[ -\frac{1}{2}\left(\frac{\Delta E + m_{0} + m_{1}E_\text{parton}}{\sigma_{0} + \sigma_{1}E_\text{parton} + \sigma_{2}\sqrt{\smash[b]{E_\text{parton}}}} \right)^{2} \right]},
\end{split}
}
\label{eq:TFexpression}
\end{equation}
where the parameters are determined independently for b jets and light-flavour jets, in three slices of $\abs{\eta_\text{parton}}$ given by $0 < \abs{\eta_\text{parton}} < 0.87$, $0.87 < \abs{\eta_\text{parton}} < 1.48$ and $1.48 < \abs{\eta_\text{parton}} < 2.5$.
In Fig.~\ref{fig:TF}, the $\Delta E$ distribution is shown for the $\Delta E = E_\text{parton} - E_\mathrm{jet}$ from simulation for all values of $E_\text{parton}$ and $\abs{\eta_\text{parton}}$.
This is compared to the $\Delta E$ distribution obtained by folding the $E_\text{parton}$ spectrum of matched partons with the transfer function.
The reasonably good agreement of the resolution and the tails of the two distributions shows that the determined transfer functions are adequate.

The disadvantage of using a LO ME is that there is no explicit treatment for final state radiation in the MEs.
As a result, the ME
does not always cover the full event information leading to a slightly reduced discrimination between both hypotheses. In addition,
background events evaluated under a $\ttbar$ hypothesis will more closely resemble the
uncorrelated hypothesis as there is no correlation between the decay products. In the
template fit part of this analysis, the small bias due to this effect is corrected for with a calibration
curve (described in Section~\ref{sec:Extractionf}), whereas in the hypothesis testing the background contribution is fixed to the predictions from
simulation, so no bias is present. \textsc{MadWeight}~\cite{Artoisenet:2010cn}, the tool used to perform the MEM likelihood calculations, can partially correct for the initial state radiation (ISR)
 effect by evaluating the LO ME at an overall partonic \pt of the $\ttbar$ system equal to the
reconstructed \pt of the system, thus properly treating five-jet events where one jet is due to ISR.
Due to final state radiation (FSR), the matching
with the LO ME, which requires four jets as input, becomes more difficult and more sensitive to systematic uncertainties related to variations on the jet energy scale or on the renormalisation/factorisation scales.
The $\ttbar$ system is reconstructed using the four selected jets based on \textsc{HitFit} in the event, the lepton and the \ptvecmiss. The \ptvecmiss quantity is assigned to the undetected neutrino from the $\ttbar$ muon+jets final state.
In the \textsc{MadWeight} likelihood calculations, every jet-quark permutation
compatible with b tagging information is taken into account.

\section{Hypothesis testing}
\label{sec:Hypothesis Testing}
The compatibility of the data with the SM hypothesis and the fully uncorrelated hypothesis is tested.
The likelihood for each event is calculated under these two hypotheses, as described in Section~\ref{sec:MEM}.
According to the Neyman--Pearson lemma, the test statistic with maximum separation power for a sample coming from either of two simple hypotheses is the likelihood ratio. This analysis uses $\lambda_\mathrm{event}$ as the discriminating variable, defined as
\begin{equation}
\lambda_\mathrm{event} = \frac{P(H_\text{uncor})}{P(H_\mathrm{cor})},
\end{equation}
where $P(H_\mathrm{cor})$ is the likelihood for the event under the SM
hypothesis and similarly $P(H_\text{uncor})$ for the uncorrelated hypothesis.

\begin{figure}
\centering
\includegraphics[width=0.45\textwidth]{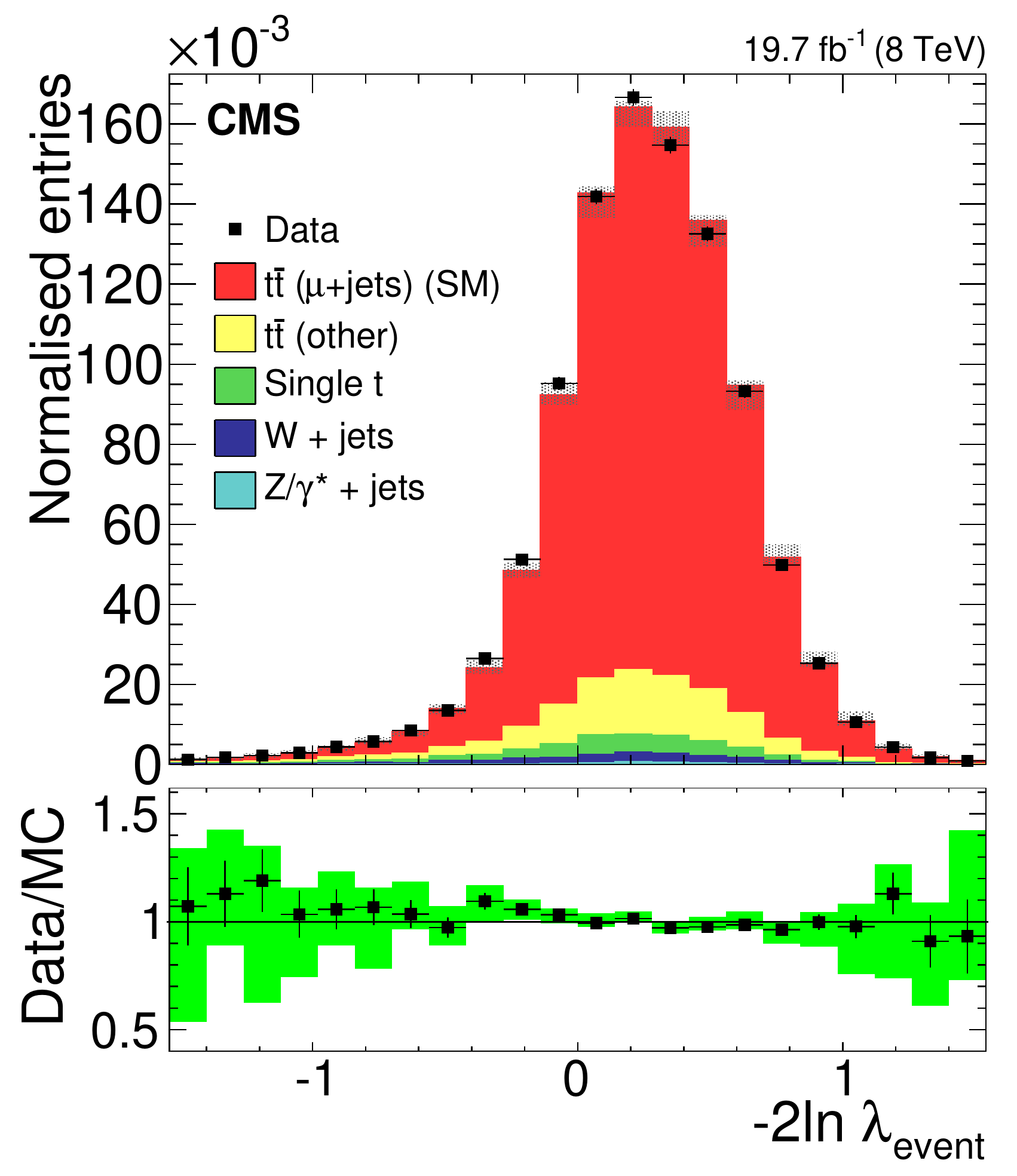}
\includegraphics[width=0.45\textwidth]{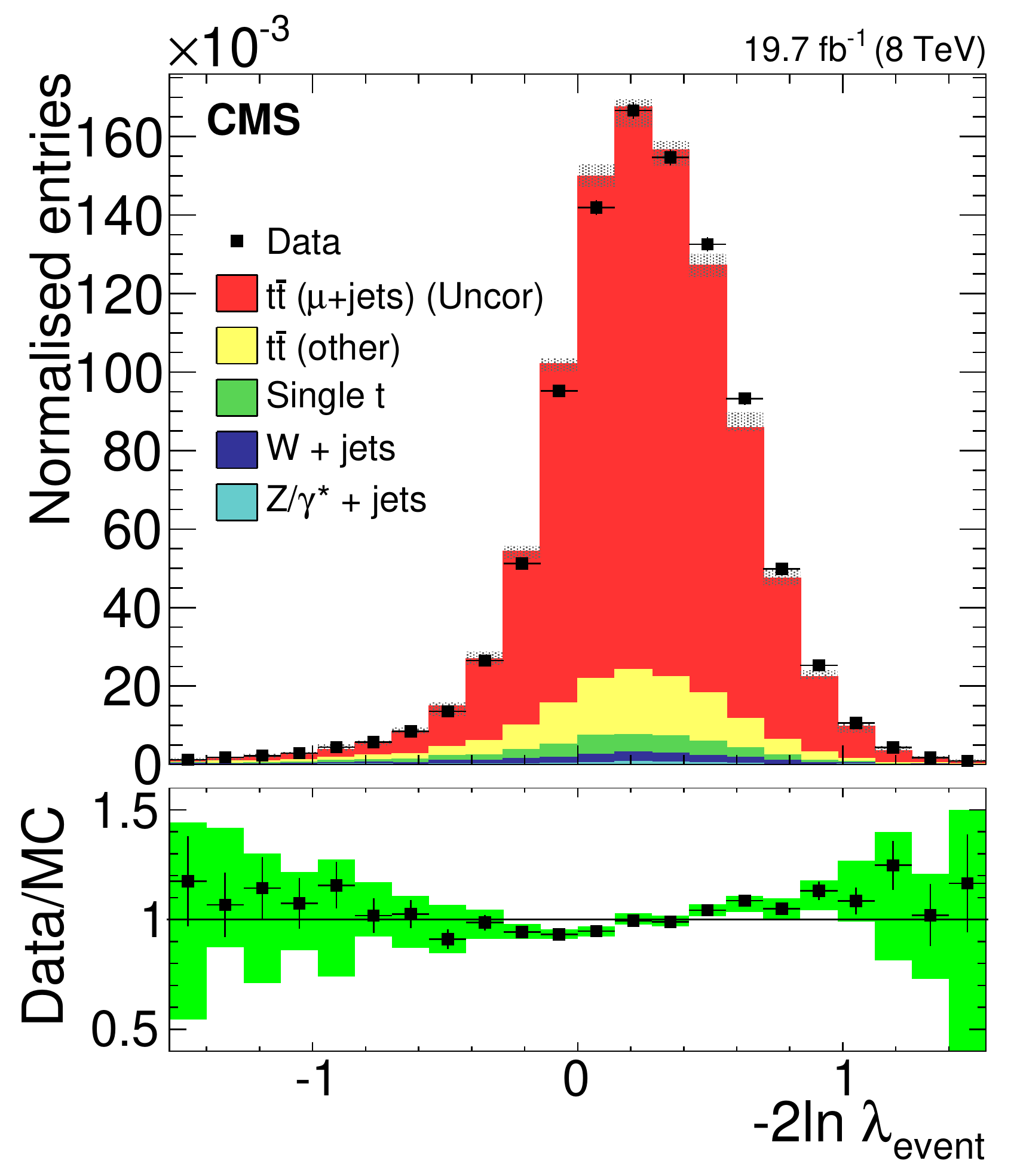}
\caption{Distribution of $-2\ln\lambda_\text{event}$. The SM $\ttbar$ simulation is used in the \cmsLeft plot and the uncorrelated $\ttbar$ simulation in the \cmsRight plot. Both data and simulation are normalised to unity. The hatched uncertainty band includes statistical and systematic uncertainties. The error bars in the ratio plot at the bottom only consider statistical uncertainties (of both data and simulation), while the uncertainty band covers both statistical and systematic uncertainties. Systematic uncertainties are described in Section~\ref{sec:Systematic Uncertainties}. The overlap of the green uncertainty band, which is constructed around the marker position, with the ratio value of 1 indicates agreement between the data and the simulation within the total uncertainty.}
\label{fig:TemplateStack}
\end{figure}

\begin{figure}[ht!]
\centering
\includegraphics[width=0.45\textwidth]{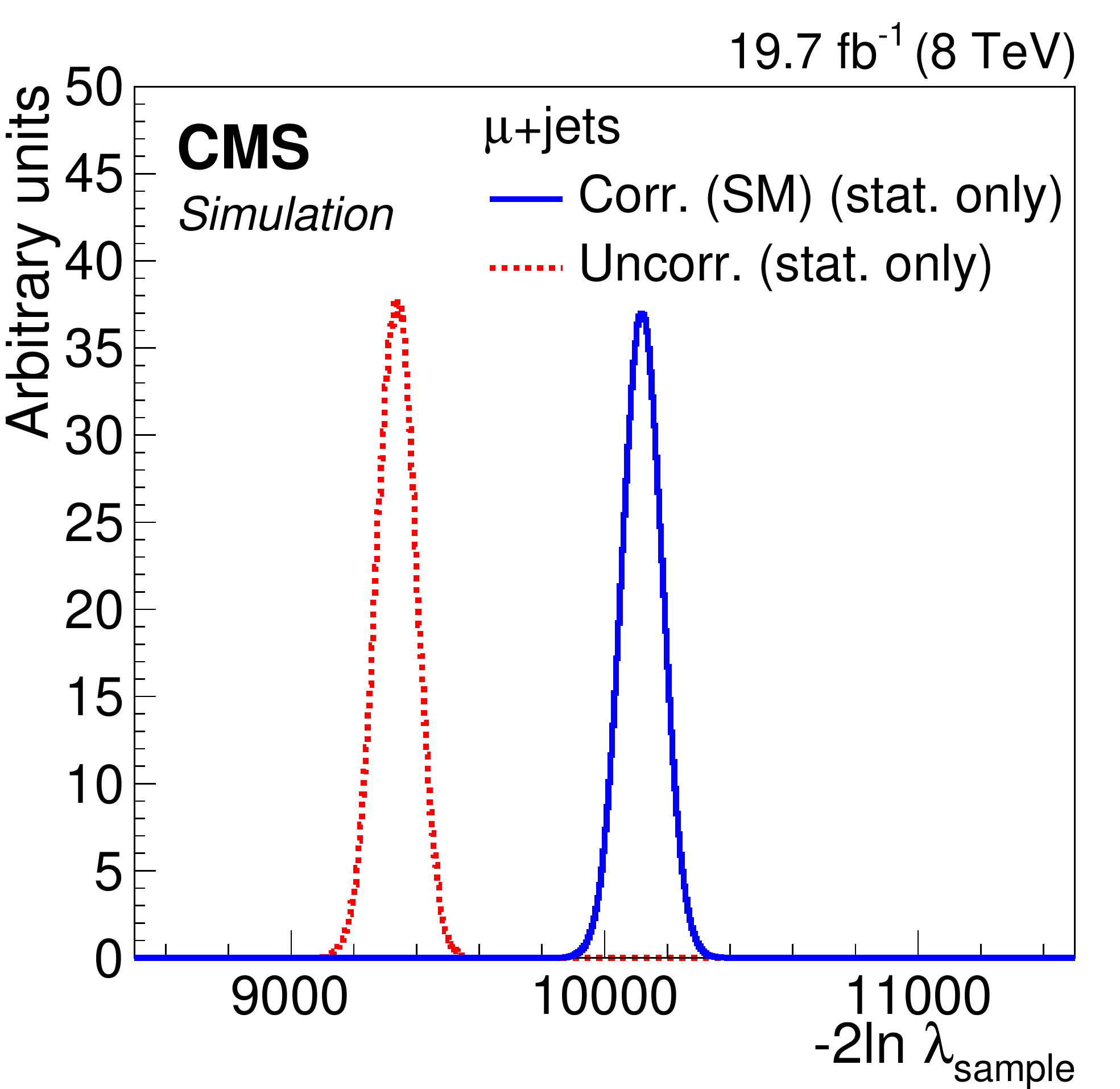}
\caption{The expected $-2\ln\lambda_\mathrm{sample}$ distribution estimated using simulation, evaluated at the data sample size. The samples in simulation contain signal and background mixed according to the theoretical cross sections, with the solid Gaussian function using SM $\ttbar$ simulation and the dashed Gaussian function using uncorrelated $\ttbar$ simulation. From this figure, the separation power can be assessed in the case when systematic effects are not considered.}
\label{fig:HypMeas}
\end{figure}

Following the prescription proposed by Cousins et al.~\cite{Cousins2005}, we use $-2\ln\lambda_\mathrm{event}$ as test statistic, a quantity hereafter referred to as the event likelihood ratio.
The distributions of $-2\ln\lambda_\mathrm{event}$ are shown in Fig.~\ref{fig:TemplateStack} for the SM $\ttbar$ sample
(Fig.~\ref{fig:TemplateStack}-\cmsLeft)
and the uncorrelated $\ttbar$ sample
(Fig.~\ref{fig:TemplateStack}-\cmsRight).
The plots show a shape comparison between data and simulation.
The differences between the SM and uncorrelated distribution are statistically significant.
The expected distribution of the sample likelihood ratio, defined as $-2\ln\lambda_\mathrm{sample} = -\Sigma2\ln\lambda_\mathrm{event}$, is calculated by drawing pseudo-experiments
with the data sample size. In the pseudo-experiments, the relative signal
and background ratios are kept fixed based on the theoretical cross sections.
These
pseudo-experiments are performed with the SM and uncorrelated event likelihood ratio distributions, respectively.
The bin width of the event likelihood ratio distribution is chosen as $0.14$ and the range of the
distribution used is $[-0.70,1.26]$. Events outside this $-2\ln\lambda_\mathrm{event}$ range of $[-0.70,1.26]$ are discarded. The shape differences of the $-2\ln\lambda_\text{event}$ distribution between the SM and uncorrelated signal hypothesis outside of this range are not statistically significant. The distribution of the sample likelihood ratios, using pseudo-experiments drawn at the data set size of $36\,800$ events within the range $-2\ln\lambda_\text{event} = [-0.70,1.26]$, is shown in Fig.~\ref{fig:HypMeas}. The solid line shows the expected Gaussian distribution of the sample likelihood ratios
for this size using the SM $\ttbar$ simulation as signal and the dashed line shows the Gaussian distribution using the uncorrelated $\ttbar$
simulation as signal. A way of quantifying the overlap between the sample likelihood ratio distributions of the two hypotheses is given by the separation power
\begin{equation}
S = \frac{\mu_{1} - \mu_{2}}{\sqrt{\alpha^{2}_{1} + \alpha^{2}_{2}}},
\label{eq.SepPower}
\end{equation}
with $\mu_{1,2}$ being the means of the distributions and $\alpha_{1,2}$ their width~\cite{Cousins2005}. The separation power is a measure for the discrimination obtainable, for the size of the data set, between the two hypotheses expressed in standard deviations ($\sigma$). Figure~\ref{fig:HypMeas} shows that a separation power of 8.8 $\sigma$ can be obtained with the MEM when only statistical effects are taken into account. The distributions will be modified by the inclusion of the systematic uncertainties described in Section~\ref{sec:Systematic Uncertainties}. The range of the $-2\ln\lambda_\text{event}$ distribution is chosen to maximise the separation power, while the binning is chosen finely enough to preserve the available separation power. In addition, the event selection (in particular the selection on the \textsc{HitFit} $\chi^2/\mathrm{ndof}$ ) has been optimised to maximize the separation power.

\section{Extraction of fraction of events with SM spin correlation}
\label{sec:Extractionf}
We extract the fraction $f$ of $\ttbar$ signal events with the SM spin correlation by performing
a template fit to the $-2\ln\lambda_\text{event}$ distribution. The fit model $M(f_\mathrm{obs}, \beta_\mathrm{obs})$ is given by
\begin{equation}
\ifthenelse{\boolean{cms@external}}
{
\begin{split}
M(f_\mathrm{obs}, \beta_\text{obs}) = & \\ (1-\beta_\text{obs} ) & \left[f_\text{obs}\, T_\mathrm{cor} + ( 1 - f_\text{obs})\, T_\text{uncor}\right] \\ & + \   \beta_\text{obs} \, T_\text{bkg},
\end{split}
}
{
M(f_\text{obs}, \beta_\text{obs}) = (1-\beta_\text{obs})\left[f_\text{obs} T_\mathrm{cor} + ( 1 - f_\text{obs}) T_\text{uncor}\right] + \beta_\text{obs} T_\text{bkg},
}
\label{eq.fitmodel}
\end{equation}
where $f_\text{obs}$ is the fraction of events with the SM spin correlation, and $\beta_\text{obs}$ is the fraction of background
in the data.
The $\ttbar$ signal SM template, the $\ttbar$ signal uncorrelated template, and the background template are denoted by
$T_\mathrm{cor}$, $T_\text{uncor}$, and $T_\text{bkg}$, respectively. The background template contains the averaged contribution of the $\ttbar$ other background with SM spin correlation and the $\ttbar$ other with no spin correlation as these contributions are the same within statistical uncertainties. Systematic uncertainties are not included in the fit model.
The parameter estimation is done using a binned maximum likelihood fit in {\sc RooFit}~\cite{RooFit2003}, using {\sc Minuit}~\cite{James:1975dr}.
The total normalisation is fixed to the observed data yield, but the relative background contribution and the fraction $f_\text{obs}$ are allowed to vary unconstrained in the fit. The binning and range of the template
distributions are fixed to those used in the hypothesis testing, where
they have been chosen to optimize the separation power between the two hypotheses.

There is a small bias in the extraction of $f_\text{obs}$ in the template fit due to the presence of background in the sample. The
background shape resembles more the behaviour of the uncorrelated template, and the size of the sample from which the background template is derived is small. The small bias is corrected for with a
calibration function. The bias is estimated from the simulation via pseudo-experiments with the observed
data set size for a range of working points ($f_\mathrm{input}$, $\beta_\mathrm{input}$). At each working point, the mean observed $f_\text{obs}$
and $\beta_\text{obs}$ are extracted to construct a 2D calibration function, used to derive $f_\mathrm{calibrated}$ as a function of the observed $f_\text{obs}$ and
$\beta_\text{obs}$. The $f_\text{obs}$- and $\beta_\text{obs}$-variables have been shifted by the weighted average of the evaluated working points to decorrelate the fit parameters.
The calibration function is given by
\begin{equation}
f = p_{0} + p_{1}\, f'_\text{obs} + p_{2}\, f'_\text{obs}\, \beta'_\text{obs},
\label{eq:FitCalibration}
\end{equation}
with $f'_\text{obs} = f_\text{obs} - 0.502$ and $\beta'_\text{obs} = \beta_\text{obs} - 0.150$.
The fit parameters of the calibration function are listed in Table~\ref{table:2DCal}.

\begin{table}[ht!]
\topcaption{Fit parameters of the 2D calibration function. The residual correlation between the fit parameters is below 10\% and is ignored.}
\renewcommand{\arraystretch}{1.1}
\centering
\begin{tabular}{ c  D{,}{\,\pm\,}{5.5} }
\hline
Parameter & \multicolumn{1}{c}{Value}\\
\hline
$p_{0}$ & 0.5004,0.0003\\
$p_{1}$ & 0.9207,0.0008\\
$p_{2}$ & -0.56,0.01\\
\hline
$\chi^{2}/\mathrm{ndof}$ & \multicolumn{1}{c}{$80/95$}\\
\hline
\end{tabular}
\label{table:2DCal}
\end{table}

\begin{figure}[ht!]
\centering
\includegraphics[width=0.45\textwidth]{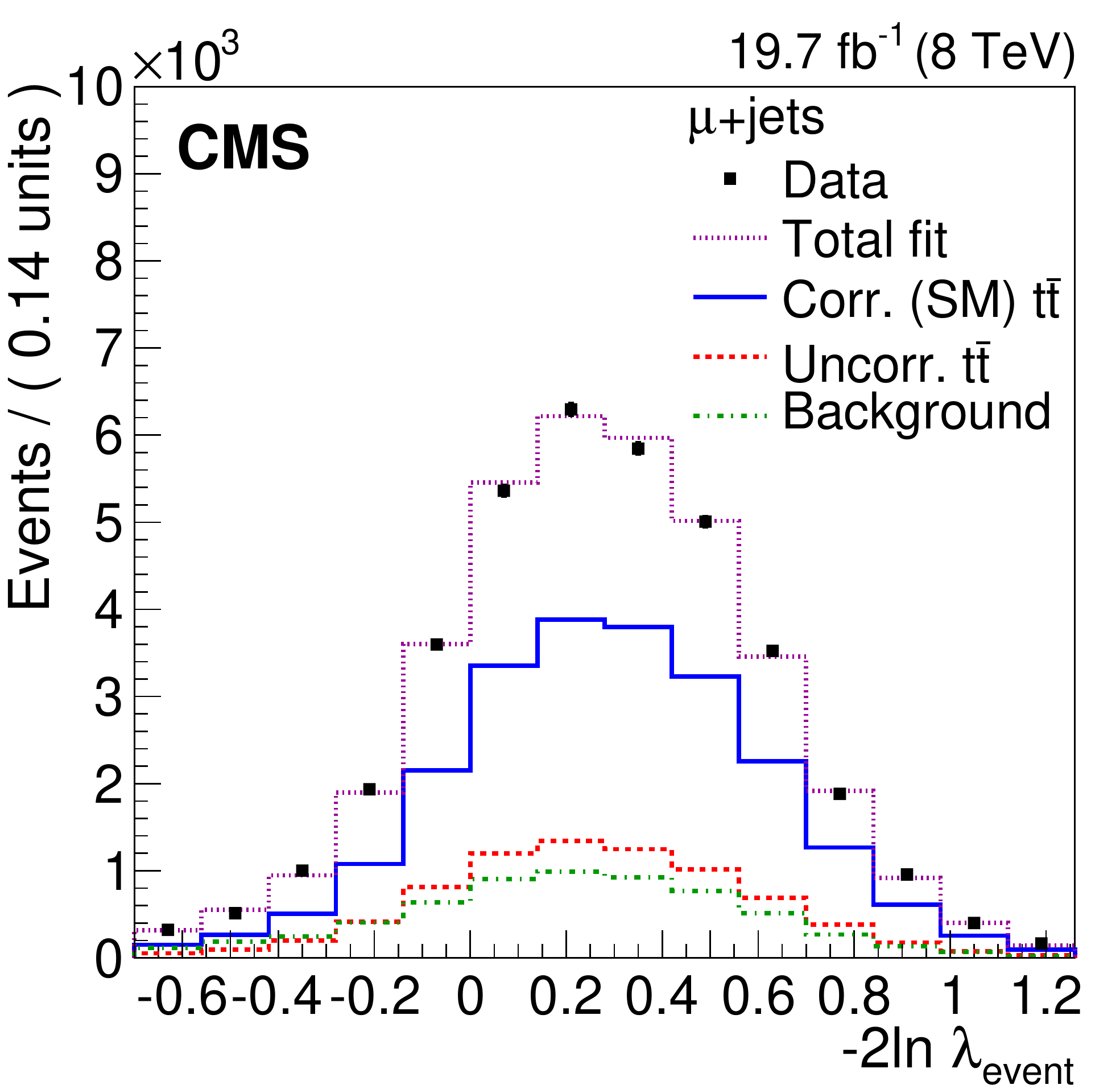}
\caption{Result of the template fit to data. The squares represent the data with the statistical uncertainty smaller than the marker size, the dotted curve is the overall result of the fit, the solid curve is the contribution of the SM signal template to the fit, the dashed curve is the contribution of the uncorrelated signal, and the dash-dot curve is the background contribution.}
\label{fig:FitNominal}
\end{figure}

It has been checked
that the initial values of the parameters in the fit model have no influence on the template fit result. The result of the template fit on
the data is shown in Fig.~\ref{fig:FitNominal} with $f_\mathrm{obs,data} = 0.747 \pm 0.092$, $\beta_\mathrm{obs,data} = 0.168 \pm 0.024$, and a $\chi^{2}/\mathrm{ndof} = 1.552$. From simulation, a background fraction $\beta$ of 15.5\% is expected in the fit range. After calibration of both the nominal result and the statistical uncertainty, the result is:
\begin{equation}
f = 0.724 \pm 0.084\stat.
\end{equation}

In the fit to the $-2\ln\lambda_\text{event}$ distribution in the range $[-0.7,1.26]$, the
correlation between $f_\mathrm{obs,data}$ and $\beta_\mathrm{obs,data}$ is around 54\%.

\section{Sources of systematic uncertainty}
\label{sec:Systematic Uncertainties}
\label{subsec:Sources of systematic uncertainty}
Systematic uncertainties affecting this analysis come from various sources, such
as detector effects, theoretical uncertainties, and mismodeling in the si\-mu\-la\-tion.
The si\-mu\-la\-tion is corrected where necessary by the use of event weights to account for efficiency differences in the data and si\-mu\-la\-tion, e.g. muon identification, isolation efficiency, trigger efficiencies, b tagging and mistagging rates and pileup modeling.
The systematic uncertainties are determined, independently of each other, by varying the efficiency correction,
resolution, or scale correction factors within their uncertainties. For some uncertainties, this is equivalent to varying the event weights,
for others, this requires recalculating the event likelihoods. In both cases, the $-2\ln\lambda_\text{event}$ distributions from which pseudo-experiments are drawn to calculate the sample likelihood ratios in si\-mu\-la\-tion or that are used as templates for the fit, are modified.
The sources of systematic
uncertainties common to the hypothesis testing and template fit are listed and explained below. The order of the list of contributions gives an indication of the relative importance of the contribution in both the template fit and the hypothesis testing. The explicit treatment of the systematic
uncertainties is explained in more detail in Section~\ref{subsec:SystHyp} for the hypothesis testing and in Section~\ref{subsec:SystF}
for the template fit.

\textbf{Limited statistical precision of si\-mu\-la\-tion:}
The $-2\ln\lambda_\text{event}$ distributions are obtained from si\-mu\-la\-tion with finite statistical precision. To estimate the effect of
the statistical precision in this distribution on the observed significance or on the template fit, each bin of the
$-2\ln\lambda_\text{event}$ distribution is varied randomly using a Poisson distribution within the statistical uncertainties. This is done
independently for each si\-mu\-la\-tion sample that contributes to the $-2\ln\lambda_\text{event}$ distribution.

\textbf{Scale uncertainty:}
SM and uncorrelated $\ttbar$ samples with varied renormalisation and factorisation scales are used to estimate the
uncertainty caused by the scale uncertainty. The renormalisation and factorisation scales are simultaneously doubled or halved with respect to their nominal values set to the sum of the transverse masses squared of the final-state particles (in the case of $\ttbar$ events this is the top quark pair and any additional parton) divided by two. The effect of the scale variation on the event selection is included.

\textbf{JES and JER effects:}
The four-momenta of all jets reconstructed in simulated events are varied simultaneously within the uncertainties of the $\pt$- and $\eta$-de\-pen\-dent JES~\cite{JES2011,JES2013} prior to the event selection.
The additional resolution correction applied to the si\-mu\-la\-tion to take into account the
resolution difference between data and si\-mu\-la\-tion is varied within the uncertainties in the si\-mu\-la\-tion. The likelihood calculations are
performed with the varied jet quantities, using the nominal transfer function. The JES uncertainty enters the measurement in two
ways: (i) acceptance effects
modify the relative contributions of the backgrounds and (ii) the event likelihood values vary due to the modified
quantities. The latter effect is dominant.

\textbf{Parton distribution functions:}
The PDF is varied within its uncertainty eigenvectors (CT10) in signal and background, and the effects are propagated through the event weights~\cite{Alekhin:2011sk, Botje:2011sn}. The procedure to propagate the effect to the $-2\ln\lambda_\text{event}$ distribution is described in~\cite{Alekhin:2011sk}.

\textbf{Top quark mass uncertainty:}
SM and uncorrelated $\ttbar$ samples with varied top quark mass values have been produced, including the
effect on the event selection. The nominal sample is simulated
with a top quark mass of 172.5\GeV, whereas the systematically varied samples are simulated with
$m_{\PQt} = 169.5\GeV$ and $m_{\PQt} = 175.5\GeV$. The $-2\ln\lambda_\text{event}$ distribution is
varied within $1/3$ of the deviation obtained with $m_{\PQt} = 175.5\GeV$ and $m_{\PQt} = 169.5\GeV$ in order to mimic the
$-2\ln\lambda_\text{event}$ variation caused by a 1\GeV uncertainty in the top quark mass world average value~\cite{ATLAS:2014wva}.

\textbf{The top quark $\boldsymbol{p_\mathrm{T}^{\PQt} }$ modeling:}
The model of $\ttbar$ production in \MADGRAPH as well as in \MCATNLO predicts a harder transverse momentum spectrum for the top quark $\pt^{t}$ than observed in the
data~\cite{PtRew1,PtRew3}. The top quark pairs might be reweighted based on the \pt spectrum of generator-level top
quarks to obtain better agreement to the measured differential cross section. This reweighting is not applied in this analysis,
but we do assign an uncertainty to the $\ttbar$ modeling by changing the event weight and propagating the effect to the $-2\ln\lambda_\text{event}$ shape.

\textbf{Background modeling and theoretical cross sections:}
We determine the relative contribution of the backgrounds using the theoretical
cross sections for the background processes. The cross sections are varied within the theoretical uncertainties~\cite{Czakon:2013} and the effects are propagated
to the analysis. The total background shape will change due to the change in relative contributions and, in the hypothesis testing, the total background fraction is fixed to the systematically varied value, whereas in the template fit, this fraction can vary freely in the fit. For the W+jets contribution, we vary the background yield by 50\% and propagate the effects to the analysis, which is ample to cover the uncertainties on the theoretical cross sections. The shape of the W+jets background template is also varied by evaluating the $-2\ln\lambda_\text{event}$ distribution without the W+jets shape included, but keeping the total background fraction fixed to the nominal value.

\textbf{Pileup:}
A $5\%$ uncertainty on the inelastic pp cross section is taken into account and propagated to the event weights~\cite{PUref}.

\textbf{The b tagging efficiency and mistag rates:}
The $\pt$- and $\eta$-dependent tagging and mistagging efficiencies for light- and heavy-flavour jets are varied within their uncertainties and are
propagated to the event weights in the si\-mu\-la\-tion~\cite{CMS-PAS-BTV}.

\textbf{Lepton trigger, identification, and isolation efficiencies:}
$\pt$- and $\eta$-dependent scale factors are applied to the si\-mu\-la\-tion to correct for efficiency differences in the data and
si\-mu\-la\-tion for the single lepton trigger, lepton identification and isolation. These scale factors are varied independently within
their uncertainties and the effects are propagated to the event weights.

The contribution of the individual systematic uncertainty sources is evaluated in the template fitting procedure described in
Section~\ref{subsec:SystF} and reported in Table~\ref{tab:systematicsfit}. The relative size of each systematic uncertainty
contribution is consistent in the hypothesis testing procedure and the template fitting.

\section{Results}
\label{sec:Results}
\subsection{Hypothesis testing}
\label{subsec:SystHyp}
To evaluate the compatibility of the data with either of the hypotheses, the systematic variations of the $-2\ln\lambda_\text{event}$ distribution
need to be propagated to the $-2\ln\lambda_\mathrm{sample}$ distribution.
We assess the effect of this event likelihood ratio fluctuation
by a Gaussian template morphing technique in which all systematic uncertainties
are evaluated simultaneously. In each pseudo-experiment, we draw a sample from the morphed template with a size equal to that of the data set,
and evaluate the sample likelihood ratio.

The $-2\ln\lambda_\text{event}$ distribution is morphed in the following way. We draw a vector $\vec{x}$ of random
numbers from a Gaussian distribution with mean 0 and width 1. Per systematic uncertainty source $k$, we have an
independent
entry $x_{k}$ in the vector. In each bin of the morphed template, the bin content $N_{i}$ is calculated as shown in
the following equation with $H(x_{k})$ a Heaviside step function and $N^\mathrm{nom}_{i}$ the original bin content:
\begin{equation}
\ifthenelse{\boolean{cms@external}}
{
\begin{split}
N_{i} = N^\mathrm{nom}_{i} +  \Sigma_{k} |x_{k}| \, & \left( \, H(x_{k}) \  \left[ N^{k,\mathrm{up}}_{i} - N^\mathrm{nom}_{i}\right] \right. \\ +  & \left. H(-x_{k}) \left[ N^{k,\mathrm{down}}_{i} - N^\mathrm{nom}_{i}\right] \, \right).
\end{split}
}
{
N_{i} = N^\mathrm{nom}_{i} + \Sigma_{k} |x_{k}| \left( H(x_{k}) \left[ N^{k,\mathrm{up}}_{i} - N^\mathrm{nom}_{i}\right] + H(-x_{k}) \left[ N^{k,\mathrm{down}}_{i} - N^\mathrm{nom}_{i}\right] \right).
}
\label{eq:TemplateMorph}
\end{equation}

Here, $N^{k,\mathrm{up}}$ and $N^{k,\mathrm{down}}$ are the bin contents of the systematically varied $-2\ln\lambda_\text{event}$ distribution for the upward and downward variation respectively. The summation runs over all systematic uncertainty sources. The systematic upward fluctuation is chosen for a
systematic source when $x_{k}$ is positive and the downward fluctuation is chosen when $x_{k}$ is negative.
This equation shows that all systematic uncertainty sources are varied simultaneously while the bin-to-bin correlations of the systematic
effect is preserved. If the systematic up- and down-effects are asymmetric in size, this
asymmetry is preserved. If the systematic up- and down-effects give a change in the same direction, the largest of the two contributions is chosen as a one-sided uncertainty while zero is used for the opposite side. Per template morphing iteration, we draw one $\vec{x}$ which gives us a varied $-2\ln\lambda_\text{event}$ distribution. From this distribution with this particular $\vec{x}$, we draw one pseudo-experiment with a size equal to that of the data set. This is done independently for the SM and uncorrelated $-2\ln\lambda_\text{event}$ distribution.

\begin{figure}[ht!]
\centering
\includegraphics[width=0.45\textwidth]{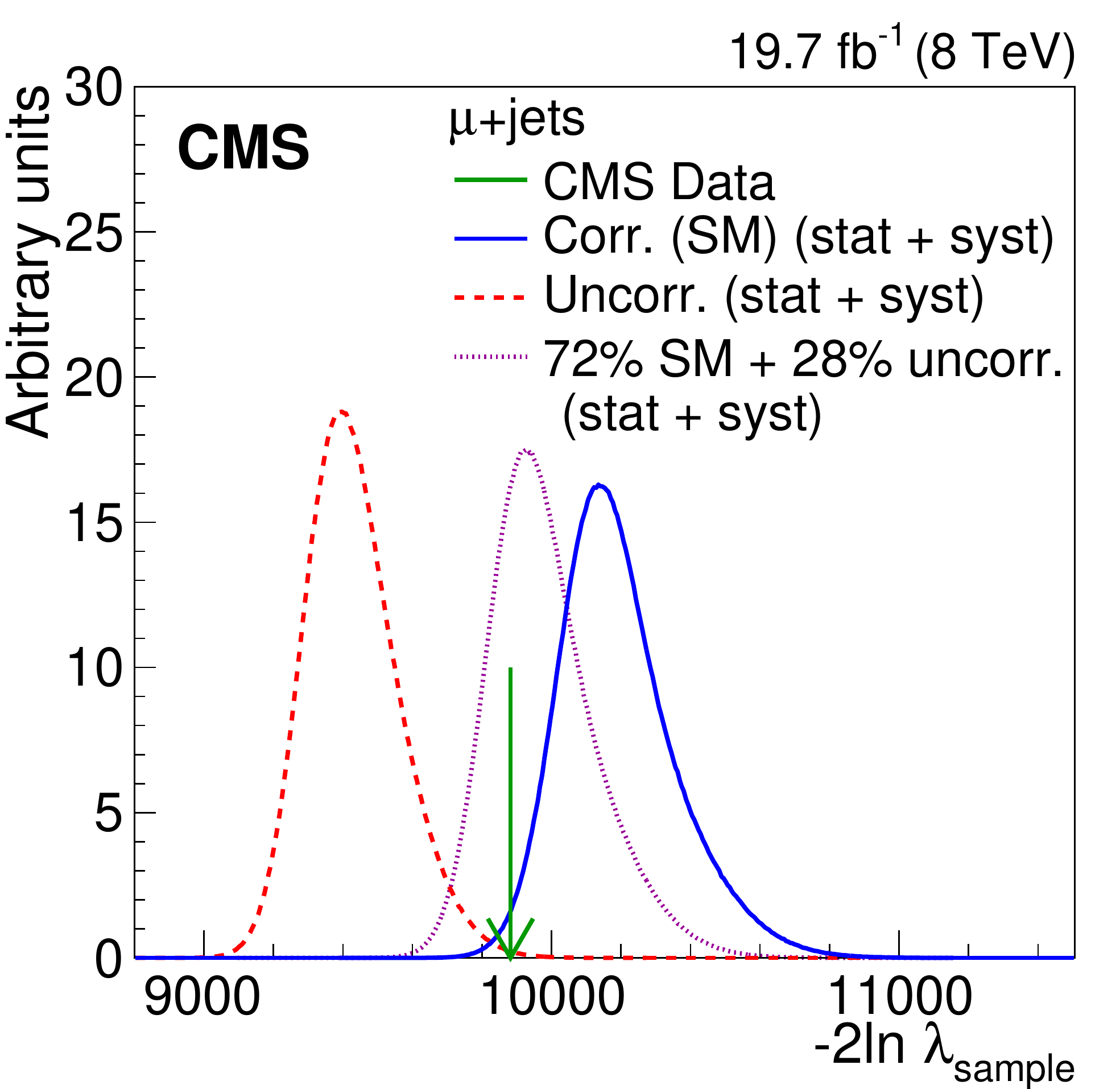}
\caption{The $-2\ln\lambda_\mathrm{sample}$ distribution in simulation, evaluated for the data set size. The samples
in simulation contain signal and background mixed according to the theoretical cross sections, with the solid distribution obtained using SM $\ttbar$ simulation and the dashed distribution obtained using uncorrelated $\ttbar$ simulation, including systematic uncertainties.
The arrow indicates the $-2\ln\lambda_\mathrm{sample}$ observed in data. The dotted curve shows a mixture of 72\% SM $\ttbar$ events and 28\%  uncorrelated $\ttbar$ events.}
\label{fig:DataTest}
\end{figure}

We perform repeated pseudo-experiments with the template morphing technique to obtain the systematically varied sample likelihood ratio
distribution shown in Fig.~\ref{fig:DataTest}. The comparison of Figs.~\ref{fig:HypMeas} and~\ref{fig:DataTest} shows the degradation of the separation power between the SM distribution and the uncorrelated distribution due to the systematic uncertainties. In addition, the result of the asymmetric behaviour of some systematic uncertainty sources is clearly visible. Performing $10^7$ pseudo-experiments is enough to po\-pu\-late the Gaussian tails in the
template morphing phase space, ensuring a smooth $-2\ln\lambda_\mathrm{sample}$ distribution with low statistical uncertainty even in the tails.

From the value of the data
sample likelihood ratio, we find that 98.7\% of the SM simulated area is above the data value,
leading to an observed agreement with the SM hypothesis of 2.2 standard deviations.
We find that 0.2\%
 of the uncorrelated simulated area is above the data value, leading to an observed agreement of
the uncorrelated hypothesis of 2.9 standard deviations. From this we can conclude that the data is more compatible
with the SM hypothesis than with the uncorrelated hypothesis. The dominant uncertainty sources are the JES, scale variation, and
the top quark mass uncertainties. The JES uncertainty is responsible for the asymmetric tails in the distribution.

As a test of the compatibility of the result in the hypothesis testing and the extraction of $f$, the hypothesis testing has been
performed with a $\ttbar$ sample constructed such that 72\% of the events contained SM correlations while the remainder 28\% had no correlation. As a result we find a sample likelihood ratio distribution, shown in Fig.~\ref{fig:DataTest}, in between the SM and uncorrelated scenario,
with a data compatibility of 0.6 standard deviations. The value measured in data, which is slightly below the mean of the distribution, is within the expectation of statistical and systematic effects. We would have achieved even better agreement had we used in simulation a value of the top quark mass equal to the world average measurement of 173.3\GeV~\cite{ATLAS:2014wva} .

\subsection{Extraction of fraction of events with SM spin correlation}
\label{subsec:SystF}
In the extraction of $f$ using a template fit to the variable $-2\ln\lambda_\text{event}$, we have the same list of
systematic uncertainty sources as described earlier, but in addition a systematic uncertainty due to the
calibration of the method is taken into account. The calibration uncertainty is obtained by propagating the uncertainties in the calibration
fit parameters shown in Table~\ref{table:2DCal} and propagating the fit uncertainty of the fit parameter $\beta_\mathrm{obs,data}$.

The systematic uncertainties are determined by fitting the data with systematically varied templates and taking the difference from the nominal fit result. The systematic contributions, taking into account the effect of the nominal calibration function, are shown in
Table~\ref{tab:systematicsfit}
where the fit uncertainty of the nominal result is also shown. The systematic uncertainty related to the finite size of the simulation samples is evaluated by fitting one pseudo-data set
in the simulation by 1000 Poisson-fluctuated templates. The Gaussian width of the fit result $f_\text{obs}$ is taken as the systematic uncertainty value. This is
done for each simulation sample independently with the uncertainties added in quadrature. In the template fit method, all systematic uncertainties
are treated as independent of each other.

\begin{table*}[ht!]
\centering
\topcaption{Sources of systematic uncertainty in the fraction $f$ of events with the SM spin correlation. There is no downward variation for the $\pt^\mathrm{t}$ modeling.}
\label{tab:systematicsfit}
\begin{tabular}{l | ..}
\hline \\ [-10pt]
Source of syst. uncer. & \multicolumn{1}{c}{Up variation} & \multicolumn{1}{c}{Down variation}\\ [+3pt]
\hline
Simulation stat. & 0.042 & -0.042\\
Scale & -0.068 & 0.124\\
JES & 0.051 & -0.090 \\
JER & -0.023 & -0.004 \\
PDF & 0.018 & 0.045\\
$m_{\PQt}$ & 0.001 & -0.034\\
top quark $\pt^{\PQt}$ modeling & 0.023 & \multicolumn{1}{c}{\NA} \\
Background modeling & 0.017 & -0.016\\
Pileup & 0.012 & -0.015\\
b tagging efficiency & -0.001  & 0.001\\
Mistag rate & 0.005 & -0.006\\
Trigger & {<}0.001 & {<}0.001\\
Lepton ID/Iso & {<}0.001 & {<}0.001\\
Calibration & 0.003 & -0.003\\
\hline \\ [-10pt]
Total syst. uncer. & \multicolumn{2}{c}{${}^{+0.15}_{-0.13}$}\\ [+3pt]
\hline
\end{tabular}
\end{table*}

The total systematic uncertainty is obtained by adding the positive and negative contributions in Table~\ref{tab:systematicsfit}
in quadrature. When both up and down systematic variations give an uncertainty in the same direction, only the largest
value is taken into account in the given direction, and no uncertainty is assigned in the opposite direction. This gives us a total systematic uncertainty of $+0.15$ and $-0.13$. The total result of the template fit
is then:
\begin{equation}
f = 0.72 \pm 0.08\stat {}^{+0.15}_{-0.13}\syst.
\end{equation}

In the assumption that there are only the SM $\ttbar$ pairs or uncorrelated $\ttbar$ pairs, this results
in an indirect extraction of $A_\text{hel}$. By making use of the relation $A^\mathrm{measured}_\text{hel} = f^\mathrm{SM}A^\mathrm{SM,MC}_\text{hel}$ where
$A^\mathrm{SM,MC}_\text{hel} = 0.324\,\pm\,0.003$ obtained in simulation, which is in good agreement with the theoretically predicted value of $A^\mathrm{SM}_\text{hel}=0.319$~\cite{BernA2004,Bern2015} which includes NLO QCD and electroweak corrections, $A^\mathrm{measured}_\text{hel} = 0.23 \pm 0.03\stat {}^{+0.05}_{-0.04}\syst $ is
obtained.
It is found that the systematic uncertainties due to JER, trigger, lepton identification and isolation efficiencies, and b tagging efficiency are not relevant compared to the statistical uncertainties associated to them. The dominant uncertainties are the JES and renormalisation/factorisation scale variation. The relative contributions of systematic uncertainty are very similar in the hypothesis testing and template fit results.

\section{Summary}
\label{sec:Summary}
The hypothesis that $\ttbar$ events are produced with correlated spins as predicted by the SM is tested using a matrix element method in the $\mu$+jets final state at $\sqrt{s} = 8\TeV$, using pp collisions corresponding to an integrated luminosity of 19.7\fbinv.
The data agree with the uncorrelated hypothesis within 2.9 standard deviations, whereas agreement with the SM is within 2.2 standard deviations. Our hypotheses are only considered up to NLO effects in the simulation, with LO matrix elements in the likelihood calculations.

Using a template fit method, the fraction of events which show SM spin correlations has been extracted. This fraction is measured to be
$f = 0.72 \pm 0.08\stat {}^{+0.15}_{-0.13}\syst$, leading to a spin correlation strength of $A^\text{measured}_\text{hel} = 0.23 \pm 0.03\stat {}^{+0.05}_{-0.04}\syst $ using the value obtained in simulation which is compatible with the theoretical prediction for $A^\mathrm{SM}_\text{hel}$ from~\cite{BernA2004,Bern2015}.
The result is the most precise
determination of this quantity in the muon+jets final state to date and is competitive with the most accurate result in the dilepton final state~\cite{AtlasResult2012}.

\begin{acknowledgments}
\label{sec:Acknowledgements}

We thank Werner Bernreuther for kindly providing the LO matrix elements describing $\ttbar$ production and subsequent decay in the $\mu$+jets channel, valid for both on- and off-shell top quarks.

We congratulate our colleagues in the CERN accelerator departments for the excellent performance of the LHC and thank the technical and administrative staffs at CERN and at other CMS institutes for their contributions to the success of the CMS effort. In addition, we gratefully acknowledge the computing centres and personnel of the Worldwide LHC Computing Grid for delivering so effectively the computing infrastructure essential to our analyses. Finally, we acknowledge the enduring support for the construction and operation of the LHC and the CMS detector provided by the following funding agencies: BMWFW and FWF (Austria); FNRS and FWO (Belgium); CNPq, CAPES, FAPERJ, and FAPESP (Brazil); MES (Bulgaria); CERN; CAS, MoST, and NSFC (China); COLCIENCIAS (Colombia); MSES and CSF (Croatia); RPF (Cyprus); MoER, ERC IUT and ERDF (Estonia); Academy of Finland, MEC, and HIP (Finland); CEA and CNRS/IN2P3 (France); BMBF, DFG, and HGF (Germany); GSRT (Greece); OTKA and NIH (Hungary); DAE and DST (India); IPM (Iran); SFI (Ireland); INFN (Italy); MSIP and NRF (Republic of Korea); LAS (Lithuania); MOE and UM (Malaysia); CINVESTAV, CONACYT, SEP, and UASLP-FAI (Mexico); MBIE (New Zealand); PAEC (Pakistan); MSHE and NSC (Poland); FCT (Portugal); JINR (Dubna); MON, RosAtom, RAS and RFBR (Russia); MESTD (Serbia); SEIDI and CPAN (Spain); Swiss Funding Agencies (Switzerland); MST (Taipei); ThEPCenter, IPST, STAR and NSTDA (Thailand); TUBITAK and TAEK (Turkey); NASU and SFFR (Ukraine); STFC (United Kingdom); DOE and NSF (USA).

Individuals have received support from the Marie-Curie programme and the European Research Council and EPLANET (European Union); the Leventis Foundation; the A. P. Sloan Foundation; the Alexander von Humboldt Foundation; the Belgian Federal Science Policy Office; the Fonds pour la Formation \`a la Recherche dans l'Industrie et dans l'Agriculture (FRIA-Belgium); the Agentschap voor Innovatie door Wetenschap en Technologie (IWT-Belgium); the Ministry of Education, Youth and Sports (MEYS) of the Czech Republic; the Council of Scientific and Industrial Research, India; the HOMING PLUS programme of the Foundation for Polish Science, cofinanced from European Union, Regional Development Fund; the OPUS programme of the National Science Centre (Poland); the Compagnia di San Paolo (Torino); MIUR project 20108T4XTM (Italy); the Thalis and Aristeia programmes cofinanced by EU-ESF and the Greek NSRF; the National Priorities Research Program by Qatar National Research Fund; the Rachadapisek Sompot Fund for Postdoctoral Fellowship, Chulalongkorn University (Thailand); and the Welch Foundation, contract C-1845.
\end{acknowledgments}

\bibliography{auto_generated}

\cleardoublepage \appendix\section{The CMS Collaboration \label{app:collab}}\begin{sloppypar}\hyphenpenalty=5000\widowpenalty=500\clubpenalty=5000\input{TOP-13-015-authorlist.tex}\end{sloppypar}
\end{document}

%% file: TOP-13-015-authorlist.tex
\textbf{Yerevan Physics Institute,  Yerevan,  Armenia}\\*[0pt]
V.~Khachatryan, A.M.~Sirunyan, A.~Tumasyan
\vskip\cmsinstskip
\textbf{Institut f\"{u}r Hochenergiephysik der OeAW,  Wien,  Austria}\\*[0pt]
W.~Adam, E.~Asilar, T.~Bergauer, J.~Brandstetter, E.~Brondolin, M.~Dragicevic, J.~Er\"{o}, M.~Flechl, M.~Friedl, R.~Fr\"{u}hwirth\cmsAuthorMark{1}, V.M.~Ghete, C.~Hartl, N.~H\"{o}rmann, J.~Hrubec, M.~Jeitler\cmsAuthorMark{1}, V.~Kn\"{u}nz, A.~K\"{o}nig, M.~Krammer\cmsAuthorMark{1}, I.~Kr\"{a}tschmer, D.~Liko, T.~Matsushita, I.~Mikulec, D.~Rabady\cmsAuthorMark{2}, B.~Rahbaran, H.~Rohringer, J.~Schieck\cmsAuthorMark{1}, R.~Sch\"{o}fbeck, J.~Strauss, W.~Treberer-Treberspurg, W.~Waltenberger, C.-E.~Wulz\cmsAuthorMark{1}
% ...
\vskip\cmsinstskip
% ...
1:~~Also at Vienna University of Technology, Vienna, Austria\\
2:~~Also at CERN, European Organization for Nuclear Research, Geneva, Switzerland\\
% ...
